\RequirePackage[mathlines]{lineno}
\documentclass[prd,twocolumn,showpacs,amsmath,amssymb]{revtex4-1}
\usepackage{overpic,graphicx}
\usepackage{dcolumn}
\usepackage{bm}
\usepackage{rotating}
\usepackage{subfigure}
\usepackage{color}

\begin{document}

\title{\bf \boldmath
Search for baryon and lepton number violating decays $D^+\to\bar\Lambda(\bar\Sigma^0)e^+$ and $D^+\to\Lambda(\Sigma^0)e^+$
}

\author{M.~Ablikim$^{1}$, M.~N.~Achasov$^{10,d}$, P.~Adlarson$^{59}$, S. ~Ahmed$^{15}$, M.~Albrecht$^{4}$, M.~Alekseev$^{58A,58C}$, A.~Amoroso$^{58A,58C}$, F.~F.~An$^{1}$, Q.~An$^{55,43}$, Y.~Bai$^{42}$
% ...
\\
\vspace{0.2cm}
(BESIII Collaboration)\\
\vspace{0.2cm} {\it
$^{1}$ Institute of High Energy Physics, Beijing 100049, People's Republic of China\\
% ...
$^{4}$ Bochum Ruhr-University, D-44780 Bochum, Germany\\
% ...
$^{10}$ G.I. Budker Institute of Nuclear Physics SB RAS (BINP), Novosibirsk 630090, Russia\\
% ...
$^{15}$ Helmholtz Institute Mainz, Johann-Joachim-Becher-Weg 45, D-55099 Mainz, Germany\\
% ...
$^{42}$ Southeast University, Nanjing 211100, People's Republic of China\\
$^{43}$ State Key Laboratory of Particle Detection and Electronics, Beijing 100049, Hefei 230026, People's Republic of China\\
% ...
$^{55}$ University of Science and Technology of China, Hefei 230026, People's Republic of China\\
% ...
$^{58}$ (A)University of Turin, I-10125, Turin, Italy; (B)University of Eastern Piedmont, I-15121, Alessandria, Italy; (C)INFN, I-10125, Turin, Italy\\
$^{59}$ Uppsala University, Box 516, SE-75120 Uppsala, Sweden\\
% ...
$^{d}$ Also at the Novosibirsk State University, Novosibirsk, 630090, Russia\\
% ...
}
}

\begin{abstract}
Using a 2.93 fb$^{-1}$ data sample of electron-positron collisions taken with the BESIII detector at a center-of-mass energy of 3.773 GeV,
which corresponds to $(8296\pm31\pm64)\times10^3 D^+D^-$ pairs, 
we search for the baryon and lepton number violating decays $D^+\to\bar\Lambda(\bar\Sigma^0)e^+$ and $D^+\to\Lambda
(\Sigma^0)e^+$. No obvious signals are found with the current statistics and upper limits on the branching
fractions of these four decays are set at the level of $10^{-6}$ at 90\% confidence level.
\end{abstract}


\maketitle

\oddsidemargin  -0.2cm
\evensidemargin -0.2cm

\section{INTRODUCTION}
\begin{figure}[htbp]\centering
\includegraphics[width=0.48\textwidth]{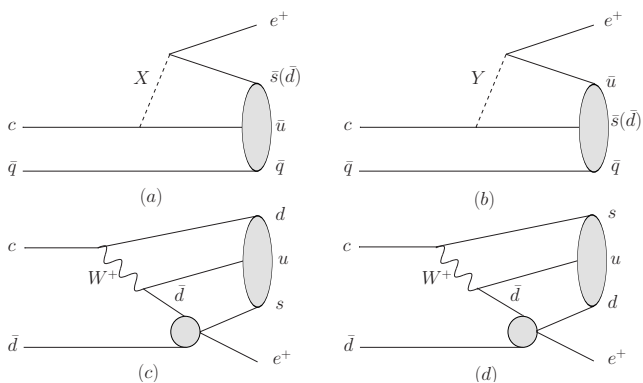}
    \caption{Feynman diagrams for the BNV decays of $D$ mesons with $\Delta(B-L)$ equal to 0 [(a) and (b)] and
    2 [(c) and (d)].}
\label{fig:feynman}
\end{figure}

    In the standard model (SM), baryon number is conserved as a consequence of the $SU(2)\times U(1)$ and
    $SU(3)$ gauge symmetries. However, the fact that there is an excess of baryons over antibaryons in
    the Universe suggests the existence of baryon number violating (BNV) processes. Thus, the search for BNV
    processes can shed light on the evolution of the Universe.
 For decades,
the decay of the proton, which is the lightest baryon, has been searched for, but no evidence for its decay has yet been found.
An alternative probe is to look for the BNV decays of heavy mesons.
Various SM extensions with BNV processes have been proposed~\cite{weinberg1979,weinberg1980,wilczek1979,weldon1980,wilczek1979b,mohapatra1980,biswal1982,hou2005}.
Under dimension six operators, BNV
processes can happen with $\Delta(B-L)=0$, where $\Delta(B-L)$ is the change in the difference between
baryon and lepton numbers. In models including heavy gauge bosons X with charge $\frac{4}{3}$ and gauge bosons Y
with charge $\frac{1}{3}$, one obtains the Feynman diagrams for BNV decays of $D$ mesons shown in
Figs.~\ref{fig:feynman}(a) and \ref{fig:feynman}(b). Another class of BNV operators is the class of dimension seven operators
    where $\Delta(B-L)=2$, as shown in the Feynman diagrams in Figs.~\ref{fig:feynman}(c) and \ref{fig:feynman}(d). Reference
\cite{wilczek1979b} argues that the decay amplitudes of these two kinds of BNV processes may be comparable. A higher
generation supersymmetry (SUSY) model predicts that the branching fraction (BF) of $D^+\to\bar\Lambda\ell^+$ is no more
than $10^{-29}$~\cite{hou2005} with the experimental limit of proton decay, where $\ell^+$ represents $e^+$ or $\mu^+$. $D^+$ BNV decays to the $\bar\Sigma^0$ baryon should have a BF at similar magnitude.
Nevertheless, an experimental search for these BNV decays will probe new physics effects
and test theoretical models beyond the SM.

    Previously, CLEO~\cite{rubin2009} and BABAR~\cite{sanzhez2011} searched for BNV processes in $D$ and $B$ decays, and recently hyperon BNV decays were searched at CLAS~\cite{McCracken:2015coa}, but no evidence
was found. The upper limits (ULs) at 90\% confidence level were set to be at the level of $10^{-5}
-10^{-8}$. 
In this paper, by analyzing 2.93 fb$^{-1}$ of data taken at a center-of-mass energy of $\sqrt{s}=3.773$ GeV
with the BESIII detector, we report the first searches for the BNV decays $D^+\to\bar\Lambda e^+$ and
$D^+\to\bar\Sigma^0 e^+$ with $\Delta(B-L)=0$, as well as $D^+\to\Lambda e^+$ and $D^+\to\Sigma^0 e^+$ with
$\Delta(B-L)=2$. 
Throughout this paper, charge-conjugated channels are implied unless explicitly stated. 

\section{THE BESIII EXPERIMENT AND DATA SAMPLE}
The BESIII detector is a magnetic spectrometer~\cite{ablikim2010} located at the Beijing Electron Positron Collider~\cite{yu2016}.
The cylindrical core of the BESIII detector consists of a helium-based multilayer drift chamber, a
plastic scintillator time-of-flight system (TOF), and a CsI(Tl) electromagnetic calorimeter (EMC), which
are all enclosed in a superconducting solenoidal magnet providing a 1.0 T magnetic field. The solenoid is
supported by an octagonal flux-return yoke with resistive plate counter muon identifier modules interleaved
with steel. The acceptance of charged particles and photons is 93\% over 4$\pi$ solid angle. The
charged particle momentum resolution at 1 GeV/$c$ is 0.5\%, and the $dE/dx$ resolution is 6\% for the
electrons from Bhabha scattering. The EMC measures photon energies with a resolution of 2.5\%~(5\%) at
1 GeV in the barrel (end-cap) region. The time resolution of the TOF barrel region is 68 ps, while that of
the end cap is 110 ps.

    Simulated samples of events produced with a {\sc geant4}-based~\cite{agostinelli2003} Monte Carlo (MC) package, which includes the geometric
description of the BESIII detector and the detector response, are used to determine the detection efficiency
and to estimate the backgrounds. The simulation includes the beam energy spread and initial state radiation
(ISR) in the $e^+e^-$ annihilations modeled with the generator {\sc kkmc}~\cite{jadach2001}. The inclusive MC samples consist
of the production of $D\bar D$ pairs including quantum coherence for all neutral $D$ modes, the non-$D\bar D$
decays of the $\psi(3770)$, the ISR production of the $J/\psi$ and $\psi(3686)$ states, and the continuum
processes incorporated in {\sc kkmc}~\cite{jadach2001}. The known decay modes are modeled with {\sc evtgen}~\cite{lange2001,ping2008}
using BFs taken from the Particle Data Group (PDG)~\cite{tanabashi2018}, and the remaining unknown decays of the
charmonium states are simulated with {\sc lundcharm}~\cite{chen2000,yang2014}. Final state radiation from charged final state particles
is incorporated with the {\sc photos}~\cite{richter1993} package.

\section{EVENT SELECTION}
To avoid possible bias, a blind analysis technique is followed where the data are viewed only after the analysis
procedure is fixed and validated with MC simulation.
The BNV decays are searched for using all tracks reconstructed within the polar angle
range $|\cos\theta|<0.93$ with respect to the beam axis. 
The $\Lambda$ and $\Sigma^0$
baryons are reconstructed via the $\Lambda\to p\pi^-$ and $\Sigma^0\to\gamma\Lambda$ decays, respectively.
Each track used to reconstruct a $\Lambda$ baryon is required to have a distance of closest approach to the
interaction point (IP) along the beam axis of less than 20 cm. 
Particle identification (PID) is applied to the
charged tracks using information from the $dE/dx$ and TOF measurements.
The confidence levels for pion, kaon, and proton hypotheses ($CL_\pi$, $CL_K$, and $CL_p$) are
calculated. The proton candidates are required to satisfy $CL_p>0.001$, $CL_p>CL_\pi$, and $CL_p>CL_K$,
while no PID is required for the pion candidates.
A vertex fit is performed to constrain the proton and pion tracks to a common vertex and the $\chi^2$ of
the fit is required to be less than 100.
The distance between the IP and the $\Lambda$ decay vertex is required to be larger than 2 standard
deviations of the vertex resolution and the invariant mass of the $p\pi^-$ combination is required to be
within $(1.110,\,1.121)$ GeV/$c^2$. 

\begin{figure}[htbp]\centering
\includegraphics[width=0.48\textwidth]{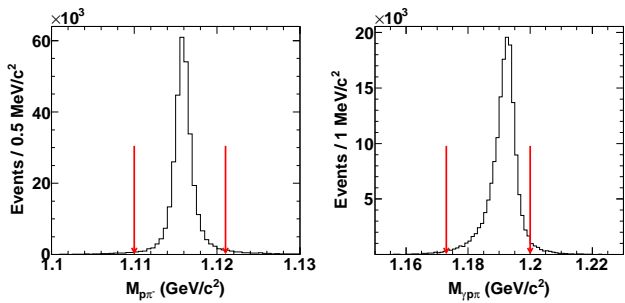}
\caption{The invariant mass distributions of the $\Lambda$ (left) and $\Sigma^0$ (right) candidates from
    the generated signal MC events, where the arrows give the mass windows.}
\label{fig:mbaryon}
\end{figure}

Photons are selected from the isolated EMC showers whose energy lost in the TOF has been recovered. The shower must
start within 700 ns of the event start time and is required to have an energy greater than 25\,(50) MeV
in the barrel (end-cap) region of the EMC. The minimum opening angle between the shower and any charged
track has to be greater than $10^\circ$. To form a $\Sigma^0$ candidate, the invariant mass of the
$\gamma p\pi^-$ combination is required to be within $(1.173,\,1.200)$ GeV/$c^2$. Figure~\ref{fig:mbaryon}
shows the invariant mass distributions of the $\Lambda$ and $\Sigma^0$ candidates in the MC simulation.

The positron candidates are required to have a distance of closest approach to the IP of less than 1 cm in
the tranverse plane and less than 10 cm along the beam axis. Positron PID is performed using $dE/dx$, TOF, and EMC information, with which the confidence levels for positron, pion, kaon, and proton
hypotheses ($CL_e$, $CL_\pi$, $CL_K$, and $CL_p$) are calculated. Positron candidates are required to
satisfy $CL_e>0.001$ and 
\begin{equation}
    \frac{CL_e}{CL_e+CL_\pi+CL_K+CL_p}>0.8.
\end{equation}
In addition, the ratio of the energy deposited in the EMC by the positron over its momentum ($E/p$) is required to be within $(0.8,\,1.2)$.

\begin{figure}[htbp]\centering
\includegraphics[width=0.48\textwidth]{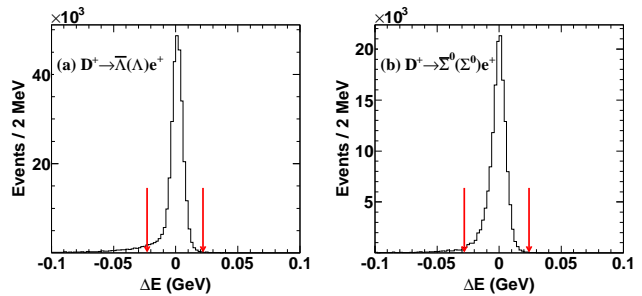}
\caption{The $\Delta E$ distributions from generated signal MC events for $D^+\to\bar\Lambda(\Lambda)e^+$ (a)
and $D^+\to\bar\Sigma^0(\Sigma^0)e^+$ (b), where the arrows give the signal windows.}
\label{fig:dlte}
\end{figure}

The BNV decays of the $D^+$ mesons are identified using the energy difference $\Delta E=E_D-E_{\rm beam}$ and the beam
constrained mass $M_{\rm BC}=\sqrt{E_{\rm beam}^2-p_D^2}$, where $E_{\rm beam}$ is the beam energy, and
$E_D$ and $p_D$ are the energy and momentum of the $D^+$ candidate in the rest frame of the $e^+e^-$ system.
When multiple candidates for a specific signal mode are present, the one with $\Delta E$ nearest to 0
is retained. The $D^+$ candidate must satisfy
$-0.023<\Delta E<0.022$ ($-0.028<\Delta E<0.024$) GeV for $D^+\to\Lambda e^+$ and $D^+\to\bar\Lambda e^+$
($D^+\to\Sigma^0 e^+$ and $D^+\to\bar\Sigma^0 e^+$), as shown in Fig.~\ref{fig:dlte}.

Studies of MC samples show that there remain a few backgrounds coming from misreconstructed $\Lambda$, e.g.,
$D^+\to\bar K^0e^+\nu_e$ and $D^+\to\bar K^{*}(892)^{0}e^+\nu_e$. However, most backgrounds are from processes
other than $\psi(3770)\to D\bar D$ where a real $\Lambda(\bar\Lambda)$ is produced. In this case, the
reconstructed positron candidates are mainly from photon conversion, decay products of pions, muons, or kaons,
as well as misidentification from other kinds of particles. Backgrounds produced due to photon conversion
are identified using the technique introduced in Ref.~\cite{xu2012}. An electron-positron pair is formed by
looping over all electrons in the event. The electron with minimum angle relative to the positron candidate is chosen. Three variables, the minimum signed distance between the electron and positron in the $xy$ plane
$\Delta_{xy}$, the polar angle of the direction of the conversion photon with respect to the vector from the IP to the
common vertex of the electron-positron pair $\theta_{eg}$, and the distance between the common vertex of
the electron-positron pair and the IP in the $xy$ plane $R_{xy}$, are defined. Events with $-2<\Delta_{xy}<1$
cm, $\cos\theta_{eg}>0.8$, and $R_{xy}>2$ are identified as background associated with photon conversions
and are rejected. To suppress backgrounds from the $e^+e^-\to q\bar q$ process and charmonium decays which may
contain a baryon-antibaryon pair, 
we require that no charged particle satisfies the proton PID criteria, except the proton from the BNV decay
candidate.

\begin{figure}[htbp]\centering
\includegraphics[width=0.48\textwidth]{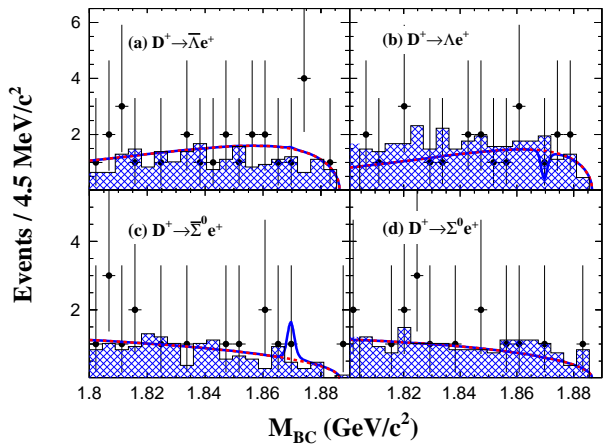}
\caption{Fits to the $M_{\rm BC}$ distributions of the accepted candidate events in data, where the dots
with error bars are data, the solid curves are the best fits, and the red dashed curves are the background
shapes. The blue hatched histograms are the MC-simulated backgrounds scaled to data according to the
luminosity.}
\label{fig:mbc}
\end{figure}

Figure~\ref{fig:mbc} shows the $M_{\rm BC}$ distributions of the accepted candidate events in data and
inclusive MC samples. A maximum likelihood fit to the $M_{\rm BC}$ distribution is performed on each distribution of data to extract the 
number of signal events in each signal decay mode. In the fit, the signal is modeled by an MC-simulated shape convolved with a Gaussian to account for the resolution difference between data and MC simulation
and the background is modeled by an ARGUS function~\cite{albrecht1990}, which has been found to be in agreement with inclusive MC
samples. The end point of the ARGUS function is fixed at the beam energy and other parameters are determined from the fit. 
The data/MC difference in $M_{\rm BC}$ resolution 
is estimated using the topologically similar decays $D^+\to K_S^0\pi^+$ and
$D^+\to K_S^0\pi^+\pi^0$ for signal channels involving $\Lambda$ and $\Sigma^0$, respectively.
The efficiencies of reconstructing the
four signal decay modes are estimated using the signal MC samples, in which signal events are generated
with unpolarized particles.

\section{SYSTEMATIC UNCERTAINTIES}
The systematic uncertainties in the searches for these BNV decays excluding those involved in the $M_{\rm BC}$
fit are summarized in Table~\ref{tab:sys}. The total number of $D^+D^-$ pairs in
the data set was previously measured in Ref.~\cite{ablikim2018} with an uncertainty of 0.9\%. The uncertainties in
the tracking and PID efficiencies of the positron are studied using $e^+e^-\to\gamma e^+e^-$ events. To
account for the difference in kinematics between the positrons in the control sample and the signal decays,
the tracking and PID efficiencies are estimated by weighting the efficiencies extracted from the control
sample in two-dimensional (momentum and $\cos\theta$) distributions. The differences of the 
weighted efficiencies between data and MC simulation, which are 0.3\% for tracking and 1.0\% for PID, are
taken as the associated systematic uncertainties. The uncertainty in the reconstruction efficiency of
$\Lambda(\bar\Lambda)$ was previously studied in Ref.~\cite{ablikim2012} using $J/\psi\to\Lambda\bar\Lambda\pi^+\pi^-$
events. The momentum-weighted difference of $\Lambda(\bar\Lambda)$ reconstruction efficiencies
between data and MC simulation is 1.5\%,
which is assigned as an uncertainty of the $\Lambda(\bar\Lambda)$
reconstruction. This includes the uncertainties in the tracking efficiencies of the pion and proton, the PID
efficiency of the proton, the decay length requirement, and the mass window. For decays involving the  
$\Sigma^0(\bar\Sigma^0)$ baryon, the uncertainty in the photon reconstruction efficiency is taken to be 1.0\%
according to the previous study using $J/\psi\to\pi^+\pi^-\pi^0$ events~\cite{ablikim2011}. The uncertainty in the
requirement of the mass window of the $\Sigma^0(\bar\Sigma^0)$ baryon is studied with $J/\psi\to pK^-\bar\Sigma^0+c.c.$
events, and is found to be negligible. The uncertainties in the $\Delta E$ requirement are estimated by smearing the MC simulated $\Delta E$ distributions with Gaussian functions
accounting for the resolution difference between data and MC simulation.
The efficiency changes after smearing are taken to be the associated
systematic uncertainties, which are 0.6\%, 0.6\%, 0.9\%, and 0.9\% for $D^+\to\bar\Lambda e^+$,
$D^+\to\Lambda e^+$, $D^+\to\bar\Sigma^0 e^+$, and $D^+\to\Sigma^0 e^+$, respectively. 
To study the uncertainty in the photon conversion
veto, we separately examine the data-MC difference in finding an extra electron in the system recoiling 
against the $D^+$ meson and that for the $R_{xy}$, $\Delta_{xy}$ and $\cos\theta_{eg}$ requirement. 
The former is studied using the selected $D^+\to K_S^0\pi^+$ vs. $D^-\to$
anything sample, and the latter with $J/\psi\to\pi^+\pi^-\pi^0,\pi^0\to\gamma e^+e^-$ events. 
Combining these two studies, we assign 0.5\% for the uncertainty in the photon conversion veto for the four
signal decay modes. The uncertainty in
the requirement of no extra proton (antiproton) is studied using the selected $D^+\to K_S^0\pi^+$ vs.
$D^-\to$ anything sample. The difference of the acceptance rates of no additional proton (antiproton) 
between data and MC simulation, 0.3\%, is assigned as the associated
uncertainty. We take 0.8\% as the uncertainty in the BF of $\Lambda\to p\pi^-$ quoted from the PDG~\cite{tanabashi2018} and
0.5\% for $\Sigma^0\to\gamma\Lambda$ by referring to a theoretical value of the BF of $\Sigma^0\to\Lambda e^+e^-$
\cite{feinberg1958}. In total, the uncertainties of the quoted BFs are 0.8\% for $D^+\to\bar\Lambda(\Lambda)e^+$ and
0.9\% for $D^+\to\bar\Sigma^0(\Sigma^0)e^+$. 
The limited MC statistics is also taken into account as a source of systematic uncertainty. 
The total systematic uncertainties are obtained by adding these uncertainties in quadrature.

\begin{table}[htbp]\centering
\caption{The systematic uncertainties excluding those involved in the $M_{\rm BC}$ fit (in \%) for the
four signal channels.}
\label{tab:sys}
\begin{tabular}{ccccc}
\hline
Source					& $\bar\Lambda e^+$ & $\Lambda e^+$ & $\bar\Sigma^0 e^+$ & $\Sigma^0 e^+$ \\\hline
$N_{D^+D^-}^{\rm tot}$ 	& 0.9				& 0.9			& 0.9			   & 0.9		  \\
$\Delta E$ cut			& 0.6				& 0.6			& 0.9			   & 0.9		  \\
$\Lambda(\bar\Lambda)$ reconstruction & 1.5	& 1.5			& 1.5			   & 1.5		  \\
$\Sigma^0(\bar\Sigma^0)$ mass window	  &     &    			& $<0.1$		   & $<0.1$		  \\
$e^+$ tracking 			& 0.3				& 0.3			& 0.3			   & 0.3		  \\
$e^+$ PID				& 1.0				& 1.0			& 1.0			   & 1.0		  \\
$\gamma$ reconstruction	& 					& 				& 1.0			   & 1.0		  \\
MC statistics			& 0.3				& 0.4			& 0.4			   & 0.4		  \\
No extra (anti-)proton		  & 0.3	& 0.3			& 0.3			   & 0.3		  \\
Photon conversion veto	& 0.5			    & 0.5		    & 0.5		       & 0.5		  \\
Quoted BF(s)			& 0.8				& 0.8			& 0.9			   & 0.9		  \\\hline
Total 					& 2.4				& 2.4			& 2.7			   & 2.7		  \\\hline
\end{tabular}
\end{table}

\section{UPPER LIMIT ESTIMATION}
\begin{figure}[htbp]\centering
\includegraphics[width=0.48\textwidth]{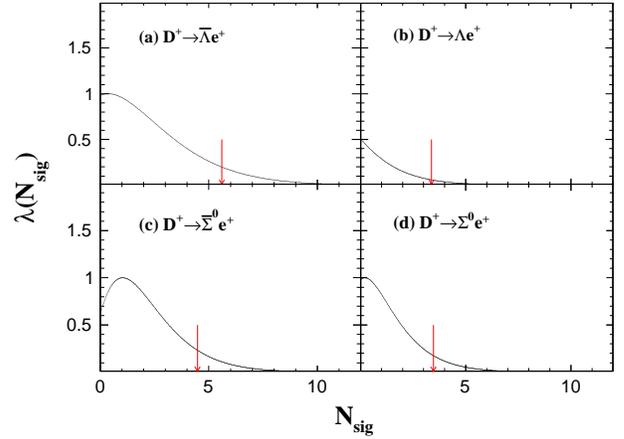}
\caption{The likelihood ratio distributions with respect to the number of signal events, where
the red arrows give the upper limits at 90\% confidence level.}
\label{fig:nsig}
\end{figure}

Since no significant signals are observed, we set the ULs on the BFs at 90\% confidence level for
the four signal decay modes. 
This is done by scanning the ratio of the likelihood value given the number of signal events and the maximum likelihood value [$\lambda(N_{\rm sig})$] in the $M_{\rm BC}$ fit.
The likelihood ratio distribution is then 
convoluted with a Gaussian function with corresponding width to incorporate the
systematic uncertainties.
The ULs on the number of signal events at 90\% confidence level ($N_{\rm sig}^{\rm UL}$)
are extracted by integrating over the physics region and finding the solution of
\begin{equation}
\int_0^{N_{\rm sig}^{\rm UL}}N_{\rm sample}dN_{\rm sig}/\int_0^\infty N_{\rm sample}dN_{\rm sig}=90\%,
\end{equation}
where $N_{\rm sample}dN_{\rm sig}$ is the number of samples with the signal events between $N_{\rm sig}$ and $N_{\rm sig}+dN_{\rm sig}$.
In addition, to account for the uncertainties introduced in the fitting method, we
vary the signal shape, background shape and fitting range and keep the maximum
$N_{\rm sig}^{\rm UL}$ given for each signal decay.
To be specific, the signal shape is varied by changing the width of the convoluted Gaussian
according to its uncertainty;
the end point of the ARGUS function is altered from 1.8865 GeV/$c^2$ to 1.8864 and 1.8866
GeV/$c^2$;
and the fit is performed within three different regions: $(1.80,\,1.89)$, $(1.81,\,1.89)$, 
and $(1.82,\,1.89)$ GeV/$c^2$.
Figure~\ref{fig:nsig} shows the likelihood ratio distributions with respect to the number of
signal events for the four signal decays.
For each signal decay mode, the UL on the BF is calculated as
\begin{equation}
\mathcal{B}^{\rm UL}=N_{\rm sig}^{\rm UL}/(2\times N_{D^+D^-}^{\rm tot}\times\varepsilon\times\mathcal{B}_{\Lambda,\Sigma^0}),
\end{equation}
where $N_{D^+D^-}^{\rm tot}$ is the total number of $D^+D^-$ pairs which was measured to be
$(8296\pm31\pm64)\times10^{3}$~\cite{ablikim2018}, $\varepsilon$ is the signal efficiency and $\mathcal{B}_{\Lambda,\Sigma^0}$
represents the BFs of the secondary decays used to reconstruct $\Lambda$ and $\Sigma^0$. Table~\ref{tab:ul}
summarizes the ULs on the numbers of signal events in data, the signal efficiencies and the corresponding
ULs on the BFs for the four signal decay modes.

\begin{table}[htbp]\centering
\caption{The ULs on the number of signal events at 90\% confidence level, the detection efficiencies not including
the BFs of the secondary decays, and the corresponding ULs on the BFs for the four signal decay modes, where the
systematic uncertainties have been included.}
\label{tab:ul}
\begin{tabular}{lccc}\hline
Mode				& $N_{\rm sig}^{\rm UL}$	& $\varepsilon$ (\%)	& $\mathcal{B}^{\rm UL}$	\\\hline
$\Lambda e^+$		& 5.6						& $31.11\pm0.14$		& $1.1\times10^{-6}$		\\
$\bar\Lambda e^+$	& 3.4						& $31.18\pm0.10$		& $6.5\times10^{-7}$		\\
$\Sigma^0 e^+$		& 4.5						& $16.31\pm0.07$		& $1.7\times10^{-6}$		\\
$\bar\Sigma^0 e^+$	& 3.5						& $16.40\pm0.07$		& $1.3\times10^{-6}$		\\\hline
\end{tabular}
\end{table}

\section{SUMMARY}
In summary, using 2.93 fb$^{-1}$ of data taken at $\sqrt{s}=3.773$ GeV with the BESIII detector, we have
searched for the BNV decays $D^+\to\bar\Lambda e^+$, $D^+\to\bar\Sigma^0 e^+$, $D^+\to\Lambda e^+$, and
$D^+\to\Sigma^0 e^+$ for the first time with the assumption of no preferred polarization of the final products. No obvious signals are found, and the ULs on the BFs of these
decays are set at 90\% confidence level, as shown in Table~\ref{tab:ul}. Our limits are far above the
prediction of the higher generation model~\cite{hou2005}.

\section*{ACKNOWLEDGEMENTS}
The BESIII collaboration thanks the staff of BEPCII and the IHEP computing center for their strong support. This work is supported in part by National Key Basic Research Program of China under Contract No. 2015CB856700; National Natural Science Foundation of China (NSFC) under Contracts Nos. 11625523, 11635010, 11735014; National Natural Science Foundation of China (NSFC) under Contract No. 11835012; the Chinese Academy of Sciences (CAS) Large-Scale Scientific Facility Program; Joint Large-Scale Scientific Facility Funds of the NSFC and CAS under Contracts Nos. U1532257, U1532258, U1732263, U1832207; CAS Key Research Program of Frontier Sciences under Contracts Nos. QYZDJ-SSW-SLH003, QYZDJ-SSW-SLH040; 100 Talents Program of CAS; INPAC and Shanghai Key Laboratory for Particle Physics and Cosmology; German Research Foundation DFG under Contract No. Collaborative Research Center CRC 1044; Istituto Nazionale di Fisica Nucleare, Italy; Koninklijke Nederlandse Akademie van Wetenschappen (KNAW) under Contract No. 530-4CDP03; Ministry of Development of Turkey under Contract No. DPT2006K-120470; National Science and Technology fund; The Knut and Alice Wallenberg Foundation (Sweden) under Contract No. 2016.0157; The Swedish Research Council; U. S. Department of Energy under Contracts Nos. DE-FG02-05ER41374, DE-SC-0010118, DE-SC-0012069; University of Groningen (RuG) and the Helmholtzzentrum fuer Schwerionenforschung GmbH (GSI), Darmstadt


\begin{thebibliography}{**}
		\bibitem{weinberg1979} S. Weinberg, Phys. Rev. Lett. {\bf 43}, 1566 (1979).
		\bibitem{weinberg1980} S. Weinberg, Phys. Rev. D {\bf 22}, 1694 (1980).
		\bibitem{wilczek1979} F. Wilczek and A. Zee, Phys. Rev. Lett. {\bf 43}, 1571 (1979).
		\bibitem{weldon1980} H. A. Weldon and A. Zee, Nucl. Phys. {\bf 173}, 269 (1980).
		\bibitem{wilczek1979b} F. Wilczek and A. Zee, Phys. Lett. B {\bf 88}, 311 (1979).
		\bibitem{mohapatra1980} R. N. Mohapatra and R. E. Marshak, Phys. Lett. B {\bf 94}, 183 (1980).
		\bibitem{biswal1982} K. Biswal, L. Maharana, and S. P. Misra, Phys. Rev. D {\bf 25}, 266 (1982).
		\bibitem{hou2005} W. S. Hou, M. Nagashima, and A. Soddu, Phys. Rev. D {\bf 72}, 095001 (2005).
		\bibitem{rubin2009} R. Rubin {\it et al.} (CLEO Collaboration), Phys. Rev. D {\bf 79}, 097101 (2009).
		\bibitem{sanzhez2011} P. del Amo Sanzhez {\it et al.} (BaBar Collaboration), Phys. Rev. D {\bf 83}, 091101 (2011).
        \bibitem{McCracken:2015coa} M. E. McCracken {\it et al.}, Phys. Rev. D {\bf 92}, 072002 (2015).
		\bibitem{ablikim2010} M. Ablikim {\it et al.} (BESIII Collaboration), Nucl. Instrum. Meth. A {\bf 614}, 345 (2010).
		\bibitem{yu2016} C. H. Yu {\it et al.}, Proceedings of IPAC2016, Busan, Korea, 2016, doi:10.18429/JACoW-IPAC2016-TUYA01.
		\bibitem{agostinelli2003} S. Agostinelli {\it et al.} ({\sc geant4} Collaboration), Nucl. Instrum. Meth. A {\bf 506}, 250 (2003).
		\bibitem{jadach2001} S. Jadach, B. F. L. Ward, and Z. Was, Phys. Rev. D {\bf 63}, 113009 (2001);
				Comput. Phys. Commun. {\bf 130}, 260 (2000).
		\bibitem{lange2001} D. J. Lange, Nucl. Instrum. Meth. A {\bf 462}, 152 (2001).
		\bibitem{ping2008} R. G. Ping, Chin. Phys. C {\bf 32}, 599 (2008).
		\bibitem{tanabashi2018} M. Tanabashi {\it et al.} (Particle Data Group), Phys. Rev. D {\bf 98}, 030001 (2018).
		\bibitem{chen2000} J. C. Chen, G. S. Huang, X. R. Qi, D. H. Zhang, and Y. S. Zhu, Phys. Rev. D {\bf 62}, 034003 (2000).
		\bibitem{yang2014} R. L. Yang, R. G. Ping, and H. Chen, Chin. Phys. Lett. {\bf 31}, 061301 (2014).
		\bibitem{richter1993} E. Richter-Was, Phys. Lett. B {\bf 303}, 163 (1993).
		\bibitem{xu2012} Z. R. Xu and K. L. He, Chin. Phys. C {\bf 36}, 742 (2012).
		\bibitem{albrecht1990} H. Albrecht {\it et al.} (ARGUS Collaboration), Phys. Lett. B {\bf 241}, 278 (1990).
		\bibitem{ablikim2018} M. Ablikim {\it et al.} (BESIII Collaboration), Chin. Phys. C {\bf 42}, 083001 (2018).
		\bibitem{ablikim2012} M. Ablikim {\it et al.} (BESIII Collaboration), Phys. Rev. D {\bf 86}, 052004 (2012).
		\bibitem{ablikim2011} M. Ablikim {\it et al.} (BESIII Collaboration), Phys. Rev. D {\bf 83}, 112005 (2011).
		\bibitem{feinberg1958} G. Feinberg, Phys. Rev. {\bf 109}, 1019 (1958).
\end{thebibliography}
\end{document}